# QUANTUM CRITICAL PHENOMENA IN LADDERS OF JOSEPHSON JUNCTIONS IN A MAGNETIC FIELD

ENZO GRANATO
*Laboratório Associado de Sensores e Materiais*
*Instituto Nacional de Pesquisas Espaciais*
*12.225 - São José dos Campos, SP Brazil*

**Abstract.** A model of a ladder of Josephson junctions in a magnetic field is considered. The topological features of the zero-temperature phase diagram, as a function of charging energy and small deviations from commensurability of the vortex lattice, are strongly dependent on the field. In addition to superconductor-insulator transitions, commensurate-incommensurate transitions and depinning by quantum fluctuations of the vortex lattice are also possible. The critical behavior of the superconductor-insulator transition at $f = 1/2$ is found to depend on the ratio between interchain and intrachain Josephson couplings and is in the universality class of the classical XY-Ising model.

## 1. Introduction

Arrays of Josephson junctions can currently be fabricated in any desired geometry and with well-controlled parameters. There has been considerable interest in this system, both experimentally and theoretically, and as a tool to investigate several interesting effects like disorder, dissipation and quantum fluctuations [1]. At low temperatures where capacitive effects dominate, the array undergoes a superconductor to insulator transition as a function of charging energy[2, 3, 4, 5, 6]. These charging effects arise from the small capacitance of the grains making up the array and leads to strong quantum fluctuations of the phase of the superconducting order parameter. The universality class of this $T = 0$ superconductor-insulator transition is currently a problem of great interest specially in relation to experiments[2, 3] and theoretical predictions of universal properties[4, 7]. In fact, some of these properties, like the universal resistance, have been corroborated by measurents[3] and numerical calculations[8]. A Josephson-junction ladder provides the simplest one-dimensional version of an array in a magnetic and is an excellent system to study several interesting zero-temperature transitions due to commensurability effects of the flux lattice introduced by the field[14, 16].

A Josephson-junction ladder can be constructed by placing superconducting islands at the sites of a one-dimensional ladder as indicated in Fig. 1. At each site $r$ of the ladder we associate a phase $\theta_r$ and charge $2en_r$, representing the superconducting element which is coupled to its neighbors by Josephson couplings $E_x$ or $E_y$. The variables $n_r$ and $\theta_r$ are conjugate to each other. If only capacitance to the ground is considered the interaction Hamiltonian is given by the self-charging model

$$H = -\frac{E_c}{2}\sum_r (\frac{d}{d\theta_r}) - \sum_{<rr'>} E_{rr'} \cos(\theta_r - \theta_{r'} - A_{rr'}) \qquad (1)$$

The first term in Eq. (1) describes quantum fluctuations induced by the charging energy $E_c = 4e^2/C$ of a non-neutral superconducting grain, where $e$ is the electronic charge and $C$ the effective capacitance. The second term is the usual Josephson-junction coupling between nearest-neighbor grains. $A_{rr'} = (2\pi/\Phi_o)\int_r^{r'} \vec{A}\cdot dr$ and $\vec{A}$ is the vector potential due to the external magnetic field and the gauge-invariant sum around a plaquette $\sum A_{rr'} = 2\pi f$ with $f = \Phi/\Phi_o$. Here $f$ is a dimensionless frustration parameter and $\Phi_o$ the flux quantum. Interesting quantum critical behavior can take place for rational values, $f = p/q$, of the frustration. Since the Hamiltonian is periodic under $f \to f + n$ ($n$ integer), $f = 0$ is equivalent to $f = p/q$ with $q = 1$.

## 2. Commensurability effects

As a result of the competition between the periodicity of the vortex lattice and the underlying pinning potential provided by the ladder, different phase transitions are possible as a function of the charging energy and the field [14]. As the magnetic field is increased from zero, a transition into a vortex state can occur where the magnetic flux first penetrates the ladder. This transition can be viewed as a commensurate-incommensurate transition described by the sine-Gordon Hamiltonian [16, 17] and is the ladder analog of flux penetration in type-II superconductors. For small fields, the phases in different branches of the ladder are locked to each other while in the vortex state exponentially interacting kinks (vortices) appear that unlock the phases. Inclusion of charging effects leads to an additional insulating phase. In the vortex state, however, the vortex lattice can also become commensurate with the ladder at rational values of the flux quanta per plaquette $f_o = p/q$. The behavior of these commensurate vortex phases as a function of small deviations from commensurability $\delta f = f - f_o$ and charging energy is an interesting problem to be addressed. One can expect that for $f = p/q$ ($p$ and $q$ relative primes) the resulting phase diagram will have a strongly dependence on $q$. In fact, many different properties of an array in a field [4, 9] depend only on $q$.

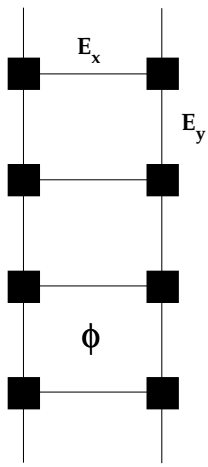

*Figure 1.* Periodic Josephson-junction ladder. Superconducting islands (squares) are coupled to their nearest neighbors with inter-($E_x$) and interchain ($E_y$) coupling constants. $\Phi$ is the magnetic flux through an elementary cell.

To study the global features of the phase diagram we use an effective action describing fluctuations from the commensurate state [14]. Starting from the path-integral representation of Eq. (1) and introducing auxiliary fields we obtain an effective action $F$ of the form

$$F = \int dx \int d\tau \{\tfrac{1}{2}K_x[(\tfrac{\delta}{\delta x}\phi_1 + \pi\delta f)^2 + (\tfrac{\delta}{\delta x}\phi_2 - \pi\delta f)^2] \\ + \tfrac{1}{2}K_\tau[(\tfrac{\delta}{\delta \tau}\phi_1)^2 + (\tfrac{\delta}{\delta \tau}\phi_2)^2] - w\cos q(\phi_1 - \phi_2)\} \quad (2)$$

where $\phi_1(x), \phi_2(x)$ are phases measuring deviations of the order parameter in each branch of the ladder from the commensurate state, and $K = K_x K_\tau \approx E_y/E_c$ and $w \approx E_x$ are effective couplings.

When $\delta f = 0$, Eq. (2) is in the form of a Gaussian approximation of two coupled classical XY models which has been studied previously with $E_c$ playing the role of an effective temperature [10, 11]. When vortices in $\phi_1, \phi_2$ are included different critical behaviors result as a function of $q$. If $q = 1$, a single Kosterlitz-Thouless (KT) transition occurs as $E_c$ increases separating a commensurate (superconducting ) phase with long-range order in the phase difference $\phi_1 - \phi_2$ and algebraic order in $\phi_1, \phi_2$, from a disordered (insulating) phase where correlations decay exponentially. For $q = 2$, recent studies indicate that it is either nonuniversal or first order if it is a single transition and it is well described by the XY-Ising model [11, 18]. For $q > \sqrt{8}$, an intermediate incommensurate (superconducting) phase is possible with algebraic order in all correlations.

The effect of small deviations from commensurability, $\delta f > 0$, can be studied by using a change of variables $\psi_1 = \phi_1 + \phi_2$, $\psi_2 = \phi_1 - \phi_2$ which leads to

$$F = \int dx \int d\tau \{ \frac{1}{4} K_x (\frac{\delta}{\delta x} \psi_1)^2 + \frac{1}{4} K_\tau (\frac{\delta}{\delta \tau} \psi_1)^2 \\ + \frac{1}{4} K_x (\frac{\delta}{\delta x} \psi_2 - 2\pi \delta f)^2 + \frac{1}{2} K_\tau (\frac{\delta}{\delta \tau} \psi_2)^2 - w \cos q\psi_2 \} \qquad (3)$$

and noting that this Hamiltonian has decoupled into a Gaussian in $\psi_1$ and a sine-Gordon in $\psi_2$ which describes the commensurate-incommensurate transition. When $\delta f$ is small, the phase difference $\psi_2$ is zero and kinks are not favorable. This implies that there is no vortices ($q = 1$) or the vortex lattice is commensurate with the ladder ($q \geq 2$). Above a critical value, $\delta f_c \approx \sqrt{w}$, kinks appear separating commensurate domains and the vortex lattice is incommensurate ($q \geq 2$). For $q = 1$ this corresponds to flux penetration in the ladder [16]. The commensurate region decreases with increasing charging energy $E_c$ and vanishes at a critical value. Including vortices in $\phi_1, \phi_2$ will result in additional disordered phases with exponentially decaying correlation functions where the ladder has an insulating behavior. For different values of $q$ the topology of the phase diagram is indicated in Fig. 2.

The topology of the phase diagrams in Fig. 2 is strongly dependent on $q$ for $f_o = p/q$, displaying direct vortex commensurate-incommensurate transitions in addition to superconductor-insulator transitions. Allowing for finite fugacities of the vortices may shrink to zero the intermediate phase at $\delta f = 0$ for $q = 3$, as renormalization-group analysis suggests [11]. However, this should persist for $q \geq 4$.

The intermediate incommensurate vortex phase at $\delta f = 0$ for $q \geq 4$ should have interesting transport properties which should be in principle experimentally measurable. In fact, the vortex lattice is effectively depinned by quantum fluctuations in this phase. If a driving current is applied transverse to the ladder, we expect that it will lead to vortex motion along the ladder and give rise to a finite resistance even at zero temperature.

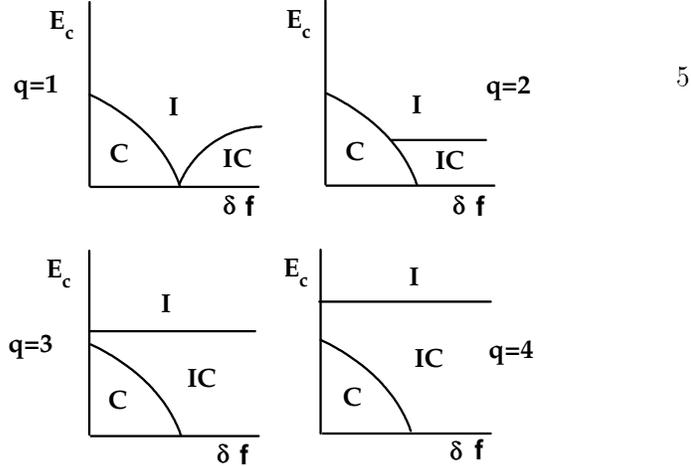

*Figure 2.* Qualitative zero-temperature phase diagram as a function of charging energy $E_c$ and small deviations $\delta f = f - f_o$ from a commensurate frustration $f_o = p/q$. C, IC and I denote commensurate-vortex, incommensurate-vortex, and insulating phases. For $q = 1$ vortices are absent in the C phase.

## 3. Critical behavior at $f = 1/2$

For increasing charging energy the one-dimensional ladder undergoes a superconductor to insulator transition at zero temperature. As discussed in Sec. 2, the details of this transition will depend both on the value of the frustration parameter $f$ and the deviation from commensurability $\delta f$. As in the case of two-dimensional superconducting films[13] and Josephson-junction arrays[4, 2, 3], the universality class of this transition is a problem of great interest.

For a chain of Josephson junctions the superconductor-insulator transition has been identified with that of the KT transition[22]. Also, for the case of two coupled chains forming a ladder, in the absence of magnetic field, this critical behavior remains unchanged. This can been seem from Eq. (3) by taking $q = 1$ and $\delta f = 0$, in which case the resulting Hamiltonian has a KT transition when vortices are included [11]. However, in the presence of the magnetic field the situation is more complicated.

For $f = 1/2$ and $\delta f = 0$, the effective Hamiltonian of Eq. (3) is expected to have a critical behavior in the universality class of the two-dimensional classical XY-Ising model [15, 18, 19] defined by the classical Hamiltonian

$$\beta H = -\sum_{rr'}[A(1 + \sigma_r\sigma_{r'}\cos(\theta_r - \theta_{r'})) + C\sigma_r\sigma_{r'}] \qquad (4)$$

The existence of both XY and Ising-like excitations in this case result from the frustration and are associated with the continuous $U(1)$ symmetry of the phases of the superconducting order parameter and the plaquette chiralities which measures the direction of circulating currents in the ladder. This model has a phase diagram consisting of three branches in the ferromagnetic region. One of them corresponds to single transitions with simultaneous loss of XY and Ising order. Further away from the branch point, this line of single transitions becomes first order. The other two lines correspond to separate KT and Ising transitions. The ladder at $f = 1/2$ is represented by a particular path trough this phase diagram which will dependend on the ratio $E_x/E_y$, between the interchain and intrachain Josephson couplings. As a consequence of this identification, the universality class of the superconductor-insulator transition should depend on the ratio between the coupling constants.

The critical behavior of the ladder at $f = 1/2$ has been studied recently using numerical techniques[15, 19]. The ladder at this value of $f$ can be regarded as a one-dimensional frustrated quantum XY (1D FQXY) model by choosing a particular gauge, $\vec{A} = xB\hat{y}$, for the vector potential in Eq. (1). In this gauge, the ladder of Fig. 1 is equivalent to an XY model with interchain coupling $E_x$ and intrachain couplings $E_y$ for one chain and $-E_y$ for the other. The critical behavior was studied using a path-integral Monte Carlo transfer matrix and finite-size scaling. In this formulation the 1D FQXY model is mapped into a 2D classical statistical mechanics problem where the ground state energy of the quantum model of finite length $L$ corresponds to the reduced free energy per unit length of the classical model defined on an infinite strip along the imaginary time direction. The parameter $\alpha = \sqrt{E_y/E_c}$ plays the role of an inverse temperature in the 2D classical model. The scaling behavior of the energy gap for kink excitations (chiral domain walls) of the 1D FQXY model corresponds to the interface free energy of an infinite strip in the classical model. For large $\alpha$ (small charging energy $E_c$), there is a gap for creation of kinks in the antiferromagnetic pattern of chiralities $\chi_p$ and the ground state has long-range chiral order. At some critical value of $\alpha$, chiral order is destroyed by quantum fluctuations with an energy gap vanishing as $|\alpha - \alpha_c|^\nu$. At the critical point, the correlation function decay as power law with an exponent $\eta$. The free energy per unit length $f(\alpha)$ on the infinite strip can be obtained from the largest eigenvalue $\lambda_o$ of the transfer matrix between different time slices as $f = -\ln\lambda_o$. To obtain $\lambda_o$, the Monte Carlo transfer-matrix method [20] was used. Similar approach has been used to study the critical behavior of the 2D frustrated classical XY model [12]. The results for the critical coupling $\alpha_c$ and critical exponents $\nu$ and $\eta$ for two different values of the ratio $E_x/E_y$ are indicated in Table 1.

TABLE 1. Critical couplings ($\alpha_c = (E_y/E_c)^{1/2}$) and exponents ($\nu$, $\eta$), obtained from finite-size scaling analysis of interfacial free energies for two values of the ratio between interchain and intrachain couplings ($E_x/E_y$)

| $E_x/E_y$ | $\alpha_c$ | $\nu$ | $\eta$ |
|---|---|---|---|
| 1 | 1.04(1) | 0.81(4) | 0.47(4) |
| 3 | 1.16(2) | 1.05(6) | 0.27(3) |

For equal couplings $E_x = E_y$, the results for the critical couplings $\eta$ and $\nu$ differ significantly from the pure 2D Ising-model exponents ($\nu = 1, \eta = 0.25$). In the XY-Ising model this correspond to a path through the phase diagram in the single transition region [18]. It is interesting to note that this result is also consistent with similar calculations for the 2D frustrated classical XY model [12]. In general, the critical behavior of a $d$ dimensional quantum model is in the same universality class of the $d+1$ dimensional classical version. This would suggest that the 1D FQXY model is the Hamiltonian limit of the 2D classical model. Apparently, this is not the case although their critical behavior appears to be in the same universality class.

For $E_x/E_y = 3$, the results indicated in Table 1 are in good agreement with pure 2D Ising values. In the phase diagram of the XY-Ising model, this correspond to a path in the decoupled region where KT and Ising transitions can take place at different points. In Fig. 3, the results for the helicity modulus are shown [19].

This quantity measures the response of the system to an imposed twist of $\pi$ along the strip and is given by $\gamma = 2\Delta F/\pi^2$ for large system sizes, where $\Delta f$ is the free-energy difference between strips with and without an additional phase mismatch. If the model were decoupled, the transition should be in the KT universality class, where one knows that the transition is associated with a universal jump of $2/\pi$ in the helicity modulus [21]. The critical coupling can be estimated as the value of $\alpha$ at which $\Delta F = \pi$ which gives $\alpha_c = 1.29$. When compared with the critical coupling for destruction of chiral order in Table 1, $\alpha_c = 1.16$, it clearly indicates that there are two decoupled transitions. Somewhere in between $\alpha = 1$ and 3 there should be a bifurcation point where a single transition with simultaneous loss of phase coherence and chiral order decouples into separate transitions. Since, the superconductor to insulator transition is to be identified with the loss of phase coherence, in the decoupled region this transition is in the KT universality class but in the single region it is in a new universality class. It is interesting to note that, unlike the zero field case, the universality class of the superconductor-insulator transition at $f = 1/2$ depends on the ratio between the inter- and intra-chain couplings.

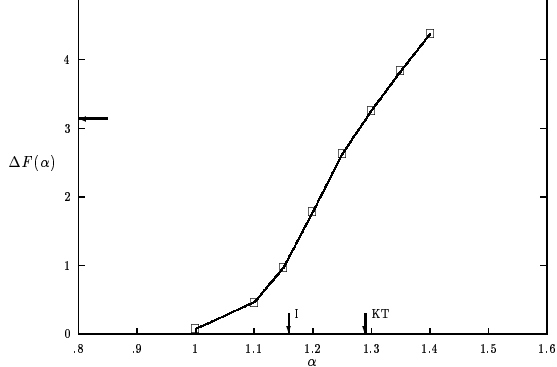

*Figure 3.* Interfacial free energy $\Delta F$ for a system of size $L = 12$ resulting from an imposed phase twist of $\pi$. Vertical arrows indicate the location of the Ising and KT transitions and the horizontal arrow the value $\Delta F = \pi$ from where the KT transition is located.

## 4. Conclusions

The zero-temperature phase diagram of a periodic Josephson-junction ladder was considered. As a function of the magnetic field, the topology of the phase diagram is strongly dependent on $q$ for $f_o = p/q$, displaying direct commensurate-incommensurate vortex transitions in addition to superconductor-insulator transitions. For $q \geq 4$, an intermediate incommensurate phase is also possible where the vortex lattice is effectively depinned by quantum fluctuations. Although these are zero-temperature properties, they can also be observed at finite but low temperatures when the thermal correlation length exceeds the sample size.

The superconductor-insulator transition at $f = 1/2$ is found to depend on the ratio between interchain and intrachain coupling constants. The result is consistent with the predictions based on the phase diagram of the classical XY-Ising model. For increasing interchain couplings the XY and

Ising-like excitations decouple, giving rise to pure Ising critical behavior for the chirality order parameter and a superconductor-insulator transition in the universality class of the KT transition. For equal couplings, there is a single transition with simultaneous loss of phase coherence and chiral order with critical exponents in agreement with the finite-temperature transition of the frustrated XY model.

## Acknowledgements

The author would like to thank the International Centre for Theoretical Physics, Trieste, Italy, for the opportunity to participate in ICTP-NATO Workshop. This work was supported by Fundação de Amparo à Pesquisa do Estado de São Paulo (FAPESP, Proc. No. 92/0963-5) and Conselho Nacional de Desenvolvimento Cientifíco e Tecnológico (CNPq).

## References

1. See articles in *Proceedings of the NATO Advanced Research Workshop on Coherence in Superconducting Networks*, Delft, 1987, Physica B **152**, 1 (1988).
2. L.J. Geerligs, M. Peters, L.E.M. de Groot, A. Verbruggen, and Mooij, Phys. Rev. Lett. **63**, 326 (1989).
3. H.S.J. van der Zant, L.J. Geerligs and J.E. Mooij, Europhysics Lett. **119** (6), 541 (1992).
4. E. Granato and J.M. Kosterlitz, Phys. Rev. Lett. **65**, 1267 (1990).
5. R. Fazio and G. Schön, Phys. Rev. B**43**, 5307 (1991).
6. E. Granato and M.A. Continentino, Phys. Rev. B **48**, 15977 (1993).
7. M.P. Fisher, G. Grinstein and S.M. Girvin, Phys. Rev. Lett. **64**, 587 (1990).
8. M.-C. Cha and S.M. Girvin, Phys. Rev. B**49**, 9794 (1994).
9. S. Teitel and C. Jayaprakash, Phys. Rev. B **27**, 598 (1983); T.C. Halsey, J. Phys. C **18**, 2437 (1985).
10. M. Yosefin and E. Domany, Phys. Rev. B**23**, 1178 (1985).
11. E. Granato and J.M. Kosterlitz, J.Phys. C **19**, L59 (1986); Phys. Rev. B **33**, 4767 (1986). J. Appl. Phys. **64**, 5636 (1988).
12. E. Granato and M.P. Nightingale, Phys. Rev. B **48**, 7438 (1993).
13. D.B. Haviland, Y. Liu and A.M. Goldman, Phys. Rev. Lett. **62**, 2180 (1989).
14. E. Granato, Phys. Rev. B **42**, 4797 (1990).
15. E. Granato, Phys. Rev. B **45**, 2557 (1992).
16. M. Kardar, Phys. Rev. B **33**, 3125 (1986).
17. F.D.M. Haldane, J. Phys. A**15**, 507 (1982).
18. E. Granato, J.M. Kosterlitz, J. Lee and M.P. Nightingale, Phys. Rev. Lett. **66**, 1090 (1991); J. Lee, E. Granato and J.M. Kosterlitz, Phys. Rev. B **44**, 4819 (1991).
19. E. Granato, Phys. Rev. B **48**, 7727 (1993).
20. M.P. Nightingale and H.W.J. Blöte, Phys. Rev. Lett. **60**, 1562 (1988); M.P. Nightingale, in *Finite Size Scaling and Numerical Simulations of Statistical Systems*, edited by V. Privman (World Scientific, Singapore, 1990).
21. D. Nelson and J.M. Kosterlitz, Phys. Rev. Lett. **39**, 1201 (1977).
22. R.M. Bradley and S. Doniach, Phys. Rev. B **30**, 1138 (1984).